\documentclass[preprint,showpacs,preprintnumbers,amsmath,amssymb,pra,superscriptaddress]{revtex4}

\usepackage[dvips]{graphicx} 
\usepackage{amsfonts}
\usepackage{amscd}
\usepackage{amsmath}    
\usepackage{enumerate}

\usepackage{amssymb}
\usepackage{amsthm}

\newtheorem{observation}{Observation}

\begin{document}

\title{Practical post-processing for quantum-key-distribution experiments}

\author{Xiongfeng Ma}
\email{xfma@iqc.ca}
\affiliation{%
Institute for Quantum Computing and Department of Physics and Astronomy, \\
University of Waterloo, 200 University Ave W., Waterloo, ON, Canada N2L 3G1 \\
}%
\author{Chi-Hang Fred Fung}
\affiliation{Department of Physics and Center of Theoretical and Computational Physics, University of Hong Kong, Pokfulam Road, Hong Kong}%
\author{Jean-Christian Boileau}
\affiliation{%
Center for Quantum Information and Quantum Control,\\
Department of Electrical \& Computer Engineering and Department of Physics, \\
University of Toronto, Toronto, Ontario, Canada, M5S 1A7 \\
}%
\author{H. F. Chau}
\affiliation{Department of Physics and Center of Theoretical and Computational Physics, University of Hong Kong, Pokfulam Road, Hong Kong}%

\begin{abstract}
Quantum key distribution (QKD) promises
unconditionally secure
key generation between two distant parties by wisely exploiting properties of quantum mechanics.
In QKD,
experimental measurements on quantum states
are transformed to a secret key and this has to be done in accordance with a security proof.
Unfortunately, many theoretical proofs are not readily implementable in experiments and do not consider all practical issues.
Therefore, in order to bridge this ``practical gap'', we integrate a few existing theoretical results together with new developments, in effect producing
a simple and complete recipe for classical post-processing that one can follow to derive a secret key from the measurement outcomes in an actual QKD experiment.
This integration is non-trivial and our consideration is both practical and comprehensive in the sense that we take into account the finiteness of the key length and consider the effects on security of several essential 
primitives
(including authentication, error handling, and privacy amplification). Furthermore, we quantify the security of the final secret key that is universally composable.
We show that the finite-size effect mainly comes from phase error estimation.
Our result is applicable to the BB84 protocol with a single or entangled photon source.

\end{abstract}

\maketitle

\section{Introduction}

Quantum key distribution (QKD)~\cite{BB_84,Ekert_91} allows two distant users to generate a secret key that is guaranteed to be unconditionally secure by the laws of quantum mechanics. Initial work on QKD has been focused on the investigation of its unconditional security 
and a few QKD protocols, such as the well-known BB84 protocol~\cite{BB_84}, have been proven to be secure in the last decade \cite{Mayers_01,LoChauQKD_99,ShorPreskill_00}.
Since then, many QKD experiments have been performed (see, e.g., references in Refs.~\cite{GRTZ_02,LoLut_review_07}).
In general, a QKD experiment involves a quantum state transmission step (where quantum states are transmitted and measured) and a classical post-processing step (where the measurement outcomes are processed classically with the help of classical communication to generate a final secret key).
Although standard security proofs (such as Ref.~\cite{ShorPreskill_00}) imply a procedure for distilling a final secret key from measurement outcomes, direct application to an {\em actual} QKD experiment is unfit.
This is because
many of these security proofs focus on the case that the key is arbitrarily long, which does not hold in practice.
It is precisely this finite-size effect that leads to reduced confidence
in
the security of the final key (mediated by the uncertainty in the post-processing tasks such as error rate estimation and error correction).
Therefore, it is imperative to quantify the finite-size effect and to provide a precise post-processing recipe that one can follow for distilling final secret keys with quantified security in real QKD experiments.
This is the purpose of this paper.
We note that, recently, lots of efforts have spent on the finite-key effect in QKD post-processing, such as Refs.~\cite{Hayashi:Finite:06,Scarani:Finite:08}.


When the key size is finite, inference on error rates and error correction can no longer be perfect as they do in the infinite-size case.
More specifically, the inferred error rates could be different from the true values and there could be leftover errors after error correction.
Consequently,
a finite-length secret key generated by a QKD system cannot be perfect
in the sense that
Alice and Bob do not share the same key and/or
Eve possesses some information about the key.
Nevertheless, the fact that the key is imperfect does not preclude it from being used in a subsequent task requiring a perfect key.
In fact,
if one can
assign 
a probability
that
the key can be regarded as an ideal one,
the use of the nonideal key as an ideal one is justified.
Indeed, this notion of security is captured by the composable security definition of QKD~\cite{BenOr:Security:05,Renner:Security:05},
which is widely adopted in the field.
QKD is composable in the sense that the final key generated is
indistinguishable from an ideal secret key except with a small failure probability.
Thus, the QKD key can be used for any subsequent cryptographic application (for instance, later rounds of QKD) requiring a perfect secret key, and the total failure probability is the sum of that of the individual composable cryptographic components.
In QKD, Alice and Bob may run a QKD system many rounds. They share a certain amount of secure key prior to each round, which can be used in the data post-processing step. The key generated by one round could be used for the next round. Composabiblity requires the key generated by all the rounds of the QKD system to be secure. In other words, an eavesdropper, Eve, knows limited amount of information about the key (if there is any) even after attacking all the rounds.

In this paper, a security definition with a failure probability (or confidence interval) is used. Our result quantifies the security of the final key generated in a QKD experiment with a failure probability, i.e., except with this probability the final key can be treated as an ideal secret key (identical and private).
This is a natural security definition for experiments and
the aforementioned composability requirement \cite{BenOr:Security:05,Renner:Security:05} is fulfilled. For instance, Alice and Bob run a QKD system $10^6$ times and keep the failure probability under $\varepsilon$ for each round. Then the total failure probability is no larger than $10^6\varepsilon$. As long as they keep $\varepsilon$ well below $10^{-6}$, the key generated in this million rounds is secure. The value of $\varepsilon$ is determined by the usage of the key in a real application.
Note that we use probability, which is more meaningful for experiments, instead of the trace distance \cite{Scarani:Finite:08}, to quantify the security.
Throughout the paper, $\varepsilon$'s with various footnotes stand for various failure probabilities.

Let us start by examining the underlying assumptions and definitions used here. We emphasize that in order to apply the scheme to a QKD system, one needs to compare these assumptions with the real setup. The assumptions used in the paper are listed as follows:
\begin{enumerate}
\item
Alice and Bob perform the BB84 protocol with a perfect single photon source or a basis-independent photon source~\cite{KoashiPreskill_03,EntanglementPDC_07}.

\item
The detection system is compatible with the squashing model \cite{Koashi_NewModel_06,TT_Thres_08,BML_Squash_08}, i.e., the input to Bob's system is assumed to be a qubit. For example, detection efficiency mismatch is not considered in this paper \cite{Mismatch_security_09}.

\item
Alice and Bob use perfect random number generators.

\item
Alice and Bob use perfect key management. 
They share a certain amount of secure key prior to running their QKD system.
\end{enumerate}

The post-processing scheme is based on a modified Shor-Prekill's security proof~\cite{ShorPreskill_00}, which is essentially Koashi's complimentary argument~\cite{Koashi_Uncer_06}. In this approach, the secure key generation is equivalent to an entanglement distillation protocol, which involves bit and phase error correction. In the post-processing, the bit error correction becomes classical error correction and the phase error correction becomes privacy amplification. We remark that our result is applicable to any physical QKD implementations that comply with the above assumptions, and it does not depend on the implementation details. 

The motivation of this paper is to give a guideline for QKD data post-processing. We start from  raw data from measurements and some pre-shared secure key bits, and produce a secret key with a quantified security definition. This can be a stepping stone for a QKD standard.  In this paper, we only present the results but not the technical details of the derivations, which will be presented in Ref.~\cite{Finite:Long:09}.

The finite key analysis is important not only from a theoretical point view, but also for experiments. For example, the efficient BB84 \cite{EffBB84_05} is proposed to increase the key generation rate. The optimal bias between the two bases, $X$ and $Z$, approaches 1 in the long key limit \cite{EffBB84_05}. In order to choose an optimal bias in the finite key case, Alice and Bob need to consider statistical fluctuations.
We remark that the proposed post-processing scheme ties up a few existing results with some new developments. Note that this integration is non-trivial and our contributions are 
as follows:
\begin{enumerate}
\item
A security definition with a failure probability is used.

\item
A strict bound for the phase error estimation is derived.

\item
An authentication scheme is applied for the error verification.

\item
The efficiency of the privacy amplification is investigated.

\item
The parameter optimization is studied.

\end{enumerate}

%
%
%
%

\section{Post-processing procedure}
Classical communication is assumed to be free in many security analyses of QKD. In practice, heavy classical communication may lead to a low key generation speed, especially for high-speed QKD setups. Moreover, some classical communication need to be authenticated (or even encrypted) in the post-processing, which means that it is not entirely free. Here, we study which part of the classical communication need to be authenticated or encrypted. For the authentication part, we rely on the LFSR-based Toeplitz matrix construction \cite{Krawczyk:Hash:94}.

The secure key used in the post-processing comes from a pre-shared secure key between Alice and Bob. For each step, we investigate the secure-key cost, $k_{xx}$, and the failure probability, $\varepsilon_{xx}$. 

The post-processing procedure is listed as follows. Note that none of following classical communication is encrypted unless otherwise stated.
\begin{enumerate}
\item
Key sift [not authenticated]: Bob discards no-click events and obtains $n$-bit raw key by randomly assigning \cite{Lutkenhaus:practical:99} the double clicks \footnote{In the case of a passive-basis-selection setup, Bob also randomly assigns basis value $X$ or $Z$ for double clicks \cite{BML_Squash_08}.}.  Note that other key sift procedures might be applied as well, see for example, Ref.~\cite{EffLoop_08}.


\item
Basis sift [authenticated]: Alice and Bob send each other $n$-bit basis information. Due to the symmetry, we can assume they pick up the same failure probability for this procedure  \cite{Krawczyk:Hash:94}
\begin{equation} \label{Finite:ReconFail}
\begin{aligned}
\varepsilon_{bs} &= n 2^{-k_{bs}+1}
\end{aligned}
\end{equation}
Here, Alice and Bob use a $2k_{bs}$-bit secure key to construct a Toeplitz matrix with a size of ($n\times k_{bs}$) by a LFSR. The authenticated tag is generated by multiplying the matrix and the message. Then they encrypt the two tags by two $k_{bs}$-bit secure keys. Since the tags are encrypted by a one-time pad, the $2k_{bs}$-bit key used for the Toeplitz matrix construction is still private \cite{Krawczyk:Hash:94}. Hence, the total secure-key cost in this step is $2k_{bs}$ and the corresponding failure probability is $2\varepsilon_{bs}$. Note that when Alice and Bob use a biased basis choice \cite{EffBB84_05}, they can exchange less than $n$-bit classical information for basis sift by data compression. Since the secure-key cost only logarithmically depends on the length of the message, we simply use $n$ for the following discussion. In the end of this step, Alice and Bob obtain $n_x$ ($n_z$)-bit sifted key in $X$ ($Z$) basis. Define the biased ratio to be $q_x\equiv n_x/(n_x+n_z)$.

\item
Error correction [not authenticated but encrypted 
\footnote{The error correction step may be done without encryption using other security proof
techniques. In this case, there may be some restriction on the error correction procedure and more
privacy amplification may be required.}]: the secure-key cost is given by
\begin{equation} \label{Finite:ECcost}
\begin{aligned}
k_{ec}=n_xf(e_{bx})H_2(e_{bx})+n_zf(e_{bz})H_2(e_{bz})
\end{aligned}
\end{equation}
where $f(x)$ is the error correction efficiency and $H_2(x)=-x\log_2(x)-(1-x)\log_2(1-x)$ is the binary entropy function. In practice, Alice and Bob only need to count the amount of classical communication used in the error correction. That is, the value of $k_{ec}$ can be directly obtained from the post-processing. After the error correction, Alice and Bob count the number of errors in $X$ ($Z$) basis: $e_{bx}n_x$ ($e_{bz}n_z$).


\item
Error verification: Alice and Bob want to make sure (with a high probability) that their keys after the error correction step are identical. Note that the idea of using error verification to replace error testing is proposed by L\"utkenhaus \cite{Lutkenhaus:practical:99}.

Comparing two procedures, authentication and error verification, one can see their commonness. In order to show the link between the two procedures, we break down the authentication procedure into two parts: Alice sends to Bob the message first and then the tag. Let us take a look at the stage that Bob just received the message sent but before the tag. Now, Alice and Bob each have a bit string. In authentication, Alice sends a tag (depending on her message) to Bob and Bob verify it. The claim of a secure authentication scheme is that if the tag pass through Bob's test, the probability that Alice and Bob share the same string is high. This can also be regarded as an error verification procedure. Hence, secure authentication schemes can be used for the error verification.

Note that the only difference between the two procedures is that in general, an authentication scheme does not care whether the tag reveals information about the message or not, but error verification does (at least for our use in QKD post-processing). This difference can be easily overcome by encrypting the tag, which has already been done in some authentication schemes including the one we are using.

Thus, in this procedure, Alice sends an encrypted tag of an authentication scheme to Bob. The cost for this step, $k_{ev}$, similar to Eq.~\eqref{Finite:ReconFail}, is
\begin{equation} \label{Finite:EVFail}
\begin{aligned}
\varepsilon_{ev} = (n_x+n_z)2^{-k_{ev}+1}
\end{aligned}.
\end{equation}

We remark that when Alice and Bob failed the error verification, they can go back to the error correction step.

\item
Phase error rate estimation [no communication]: random sampling. Alice and Bob can estimate the phase error rates in $X$ and $Z$ bases, $e_{px}$ and $e_{pz}$ separately. Take $Z$ basis for example. The probability of $e_{pz}>e_{bx}+\theta_x$ is $P_{\theta x}$ \cite{Finite:Long:09}
\begin{equation} \label{Finite:PhFailZ}
\begin{aligned}
P_{\theta x} &<
\frac{\sqrt{n_x+n_z}}{\sqrt{n_xn_ze_{bx}(1-e_{bx})}} 2^{ -(n_x+n_z)\xi_x(\theta_x) } \\
\end{aligned},
\end{equation}
where the $\xi_x(\theta_x)$ is defined by $\xi_x(\theta_x) \equiv H_2(e_{bx}+\theta_x-q_x\theta_x)-q_xH_2(e_{bx})-(1-q_x)H_2(e_{bx}+\theta_x)$
with $q_x=n_x/(n_x+n_z)$. A similar formula for $P_{\theta z}$ can also be derived. Then the total failure probability of phase error rate estimation, $\varepsilon_{ph}$, is given by
\begin{equation} \label{Finite:PhFail}
\begin{aligned}
\varepsilon_{ph} &\le P_{\theta x}+P_{\theta z}
\end{aligned}.
\end{equation}
In a highly non-likely case when $e_{bx}=0$ ($e_{bz}=0$), one can replace it by $n_xe_{bx}=1$ ($n_ze_{bz}=1$) to get around the singularity \cite{Finite:Long:09}. One can see that $\xi_x(\theta_x)$ is positive when $\theta_x>0$ and $0\le e_{bx}, e_{bx}+\theta_x \le1$, due to concavity of the binary entropy function $H_2(x)$. Note that in the limit of a large $n$, $\theta$ can be chosen small. In this case, Eq.~\eqref{Finite:PhFail} yields a similar result used in the literature, such as Refs.~\cite{ShorPreskill_00,EntanglementPDC_07}.

\item
Privacy amplification [authenticated]: Alice generates an $(n_x+n_z+l-1)$-bit random bit string and send to Bob through an authenticated channel. Alice and Bob use this random bit string to generate a Toeplitz matrix. The final key (with a size of $l$) will be the product of this matrix (with a size of $(n_x+n_z)\times l$) and the key string (with a size of $n_x+n_z$) after passing through the error verification. 
The failure probability of the privacy amplification is given by
\begin{equation} \label{Finite:PAFail}
\begin{aligned}
\varepsilon_{pa} &= (n_x+n_z+l-1) 2^{-k_{pa}+1} + 2^{-t_{oe}}
\end{aligned},
\end{equation}
where $k_{pa}$ is the secure-key cost for the authentication and $t_{oe}$ is defined by
\begin{equation} \label{Finite:PAkey}
\begin{aligned}
l &= n_x[1-H_2(e_{bz}+\theta_z)] \\
&~~~+n_z[1-H_2(e_{bx}+\theta_x)]-t_{oe} \:. \\
\end{aligned}
\end{equation}
The first term in Eq.~\eqref{Finite:PAFail} gives the failure probability of the authentication for the $(n_x+n_z+l-1)$-bit random bit string transmission. The second term in Eq.~\eqref{Finite:PAFail} gives the failure probability of the privacy amplification given the Toeplitz matrix \footnote{In the equivalent entanglement distillation protocol used for the security proof \cite{ShorPreskill_00,Koashi_Uncer_06}, the second term in Eq.~\eqref{Finite:PAFail} gives the failure probability of the phase error correction.}.

\item
The final secure key length (net growth~\footnote{Since QKD is a key expansion process, it requires some pre-shared secret bits to start with and thus they have to be accounted for when calculating the final key length.}) is given by
\begin{equation} \label{Finite:KeyFinal}
\begin{aligned}
NR &\ge l-2k_{bs}-k_{ec}-k_{ev}-k_{pa} \\
\end{aligned}
\end{equation}
with a failure probability of
\begin{equation} \label{Finite:FailFinal}
\begin{aligned}
\varepsilon\le2\varepsilon_{bs}+\varepsilon_{ev}+\varepsilon_{ph}+\varepsilon_{pa} \\
\end{aligned},
\end{equation}
where $l$ is given by Eq.~\eqref{Finite:PAkey}.

\end{enumerate}

One can see that when $n_x+n_z\gg 2k_{bs}+k_{ev}+k_{pa}+t_{oe}$, the final key length given by Eq.~\eqref{Finite:KeyFinal} is essentially the same to the one given by the Shor-Preskill's proof \cite{ShorPreskill_00}.


\section{Parameter optimization}
In order to maximize the final secure key length in the post-processing, Alice and Bob need to consider the failure probabilities from all steps and the corresponding secure-key costs. That is, they need to optimize the key rate, Eq.~\eqref{Finite:KeyFinal}, subject to Eq.~\eqref{Finite:FailFinal}. The parameters to be optimized are: biased ratio $q_x$, various secure-key costs ($k_{bs}$, $k_{ec}$,  $k_{ev}$, $k_{pa}$, $t_{oe}$) and security parameters ($\varepsilon_{bs}$, $\varepsilon_{ev}$, $\varepsilon_{ph}$, $\varepsilon_{pa}$).

In practice, Alice and Bob can calibrate the QKD system to get an estimate of the transmittance $\eta$, the error rates $e_{bx}$ and $e_{bz}$. Through some rough calculation of the target length of the final key, they decide the acceptable confidence interval $1-\varepsilon$ and fix the length of the experiment, $N$, the number pulses sent by Alice. Then roughly, the length of the raw key is $n=N\eta$. Thus, in the optimization procedure, the given values (constraints) are $\varepsilon$, $n$, $e_{bx}$ and $e_{bz}$.

The failure probability $\varepsilon$ is chosen by Alice and Bob according to later practical use of the final key. This relates to the aforementioned composability requirement \cite{BenOr:Security:05,Renner:Security:05}. For instance, Alice and Bob plan to use the QKD system for a million times, and set $\varepsilon$ for each round. Then the total failure probability for this one-million-round use is $10^6\varepsilon$, which should be below some threshold depending on the message security level. From here, one can see that the choice of $\varepsilon$ is not strictly pre-determined. That is, the final security parameter, $\varepsilon$, can slightly deviate from the pre-determined one.

Denote the probability for Alice and Bob to choose $X$ basis to be $p_x$. After the basis sift, Alice and Bob share an $n_x$-bit ($n_z$-bit) key in $X$ ($Z$) basis, where roughly (due to fluctuations) $n_x\approx p_x^2n$ and $n_z\approx(1-p_x)^2n$. Thus the biased ratio is given by $q_x\approx p_x^2/[p_x^2+(1-p_x)^2]$. In a realistic case, Alice and Bob can optimize $p_x$ first, and then optimize other parameters after the error verification part when the real values of $n_x$, $n_z$ $e_{bx}$ and $e_{bz}$ are fixed (known). In this procedure, the biased ratio cannot be strictly optimized due to fluctuations and calibration errors, while other parameters can be well optimized. 
In the end, they obtain a secure key rate and calculate the total failure probability with these parameters.



The error correction and phase error rate estimation mainly depend on the biased ratio. Thus, Alice and Bob can group the failure probabilities and secure key costs into two parts 
by defining $\varepsilon_3\equiv2\varepsilon_{bs}+\varepsilon_{ev}+\varepsilon_{pa}$ and $k_3\equiv 2k_{bs}+k_{ev}+k_{pa}+t_{oe}$, see Eqs.~\eqref{Finite:PAkey}, \eqref{Finite:KeyFinal} and \eqref{Finite:FailFinal}. The final secure key length can be rewritten as
\begin{equation} \label{Finite:KeyOpt}
\begin{aligned}
NR &\ge n_x[1-f(e_{bx})H_2(e_{bx})-H_2(e_{bz}+\theta_z)] \\
&~~~+n_z[1-f(e_{bz})H_2(e_{bz})-H_2(e_{bx}+\theta_x)]-k_3 . \\
\end{aligned}
\end{equation}
We remark that if the contribution from one basis is negative in Eq.~\eqref{Finite:KeyOpt}, Alice and Bob should use the detections from this basis for the parameter estimation only, but not the key generation.

The optimized secure-key cost for each step is given by \cite{Finite:Long:09}
\begin{equation} \label{Finite:Opt3k}
\begin{aligned}
t_{oe} &= \frac{k_3}{5}-\frac45-\frac1{5}\log_2A \\
k_{bs} 
&= t_{oe}+1+\log_2n \\
k_{ev} 
&= t_{oe}+1+\log_2(n_x+n_z) \\
k_{pa} 
&= t_{oe}+1+\log_2(n_x+n_z+l-1) ,\\
\end{aligned}
\end{equation}
where $A=n^2(n_x+n_z)(n_x+n_z+l-1)$. The corresponding failure probability is
\begin{equation} \label{Finite:Opt3e}
\begin{aligned}
\varepsilon_{3} = 5A^{1/5}2^{-(k_3-4)/5} \\
\end{aligned}.
\end{equation}

When the final key length is much larger than 37 bits, Alice and Bob can set
\begin{equation} \label{Finite:kmax}
\begin{aligned}
k_3 = -5\log_2\varepsilon+4\log_2n+50 \\
\end{aligned}
\end{equation}
and the failure probability is $\varepsilon_3<10^{-2}\varepsilon$. Since Alice and Bob will recalculate the failure probability in the end and allow the final $\varepsilon$ having small deviations from the pre-determined value, they can safely use $\varepsilon_{ph}=\varepsilon$ in the optimization. Thus, the simplified optimization problem only has three parameters to be optimized: $q_x$, $\theta_x$ and $\theta_z$, given $\varepsilon_{ph}=\varepsilon-\varepsilon_{3}\approx\varepsilon$.

\begin{observation} \label{Obs:MainPhase}
The main effect of the finite key analysis for the QKD post-processing stems from the phase error rate estimation.
Inefficiencies due to authentication, bit error correction, and privacy amplification are relatively insignificant.
\end{observation}

This can be easily seen from Eqs.~\eqref{Finite:Opt3e} and \eqref{Finite:kmax}. Even in an extreme case that $\varepsilon=10^{-30}$ and $n=10^{30}$, the secure key cost of all the parts other than the phase error rate estimation, given by Eq.~\eqref{Finite:kmax}, is 947 bits ($\ll n$) and its corresponding failure probability $\varepsilon_3<10^{-32}$.

\section{An example}
Now let us consider an example of the post-processing. Suppose $N=10^{10}$, $\eta=10^{-3}$, (then $n\approx N\eta=10^7$), $e_{bx}=e_{bz}=4\%$ and $\varepsilon=10^{-7}$. It is not hard to see that the final key length is much larger than 30 bits. Thus, we can use Eq.~\eqref{Finite:kmax} to calculate the secure-key cost, $k_3=202$ bit.


Now the problem becomes: given $n=10^7$, $e_{bx}=e_{bz}=4\%$ and $\varepsilon=10^{-7}$, optimize the parameters: $\theta_x$, $\theta_z$ and $q_x$. Through a numerical program, we get $\theta_x=1.07\%$, $\theta_z=0.84\%$ and $q_x=99.8\%$ (or $p_x=96.0\%$). Note that, in this case, the bases $X$ and $Z$ are interchangeable due to the symmetry.

With these parameters and Eq.~\eqref{Finite:Opt3k}, the final secure key length is $4.41$ Mb and its corresponding security parameter is $\varepsilon=1.0095\times10^{-7}$ (very close to the predetermined value $10^{-7}$).


In the simulation, we assume the error correction efficiency is 100\% (the Shannon limit). In this case, the difference between the ``asymptotic-key" length ($5.15$ Mb) and the ``finite-key" length ($4.41$ Mb) 
comes from the finite statistical analysis. Note that all the rest cost is only $k_3=259$ bit and $\varepsilon_3=9.5\times10^{-10}$. This is consistent with Observation \ref{Obs:MainPhase}: the cost (and the failure probability) due to the finite key analysis mainly comes from the phase error rate estimation.


\section{Further discussion}
\begin{enumerate}
\item
In the privacy amplification step, Alice and Bob need a common matrix to generate the final secure key. The current way to construct the matrix is by Alice sending a random bit string to Bob, which requires authenticated classical communication. An alternative way is by each of them generating a matrix with a pre-shared secret key. The main advantage of the second method is that no classical communication is needed for the privacy amplification part. In this case, the error verification step can be done before or after the privacy amplification.

From the LFSR-based Toeplitz matrix construction, we know that Toeplitz matrices can be generated by a much shorter random string \cite{Krawczyk:Hash:95}. By consuming a $k_{pa}$-bit secure key, Alice and Bob construct a LFSR-based Toeplitz matrix with a size of $(n_x+n_z)\times l$, where $l$ the key length after the privacy amplification. There are two related quantities need to be investigated here: the value of $l$ and its corresponding failure probability, $\varepsilon_{pa}$.

\item
In the security proof, we assume the detection system is compatible with the squashing model, where the single-mode assumption is used, and the imperfection of $X$ and $Z$ measurements and efficiency mismatch are not considered \cite{BML_Squash_08}. It is interesting to consider the detector efficiency mismatch with the finite key analysis \cite{Mismatch_security_09}.

\item
As noted in Ref.~\cite{Practical_05}, the finite-key analysis for the decoy-state QKD is a hard problem. In the decoy-state QKD, the fluctuation comes from not only statistics but also hardware imperfections. The question is where the main contribution of the fluctuation comes from and how to quantify these fluctuations. Since QKD systems with coherent states are most widely used in experiments, investigating the finite key effect in decoy-state QKD is an important step towards a QKD standard.

\item
In order to compare our finite-key analysis to others, such as the one by Scarani and Renner \cite{Scarani:Finite:08}, one has to make sure the underlying assumptions (definitions) are the same. Note that in Scarani and Renner's analysis, a trace distance is used as for the security definition. For example, it is interesting to investigate how to quantify the efficiency of authentication with the trace distance.
\end{enumerate}

\section{Acknowledgments}
We thank C.~Erven, N.~Godbout, M.~Hayashi, D.~W.~Leung, H.-K.~Lo, N.~L\"utkenhaus, M.~Koashi, X.~Mo, B.~Qi, R.~Renner, V.~Scarani, D.~Stebila, K.~Tamaki, W.~Tittel, Q.~Wang, Y.~Zhao and other participants to the workshop \emph{Quantum Works QKD Meeting (Waterloo, Canada)} and \emph{Finite Size Effects in QKD (Singapore)} for enlightening discussions. X.~Ma especially thanks H.~F.~Chau for hospitality and support during his visit at the University of Hong Kong. This work is supported from the NSERC Innovation Platform Quantum Works, the NSERC Discovery grant, the FWF (START prize), the University of Waterloo, the RGC grant No.~HKU 701007P of the HKSAR Government and the Postdoctoral Fellowship program of NSERC.

\bibliographystyle{apsrev}

\bibliography{Bibli}


\end{document}